# A new approach to business fluctuations: heterogeneous interacting agents, scaling laws and financial fragility

Domenico Delli Gatti,[a] Corrado Di Guilmi,[b] Edoardo Gaffeo,[c]
Gianfranco Giulioni,[b] Mauro Gallegati,[b*] Antonio Palestrini,[b]

[a] *Institute of Quantitative Methods and Economic Theory, Catholic University of Milan,
Largo Gemelli1, I-20123 Milan, Italy*
[b] *Department of Economics, Università Politecnica delle Marche,
Piaz.le Martelli 8, I-60121 Ancona, Italy*
[c] *Department of Economics and CEEL, University of Trento,
Via Inama 5, I-33100 Trento, Italy*

September 6[th], 2003

**Abstract**

In this paper we discuss a scaling approach to business fluctuations. Our starting point consists in recognizing that concepts and methods derived from physics have allowed economists to (re)discover a set of stylized facts which have to be satisfactorily accounted for in their models. Standard macroeconomics, based on a reductionist approach centered on the representative agent, is definitely badly equipped for this task. On the contrary, we show that a simple financial fragility agent-based model, based on complex interactions of heterogeneous agents, is able to replicate a large number of scaling type stylized facts with a remarkable high degree of statistical precision.

---

[*] Corresponding author. E-mail: gallegati@dea.unian.it (M. Gallegati).

# 1. Introduction

Recent explorations in industrial dynamics have detected two empirical regularities which are so widespread across countries and persistent over time to be characterized as universal laws: (*i*) the distribution of firms' size is right skew and can be described by a Zipf or power law probability density function (Axtell, 2001; Gaffeo *et al.* 2003); (*ii*) the growth rates of firms' output and countries' GDP follow a Laplace distribution (Stanley *et al.*, 1996; Amaral *et al.*, 1997).

The fact that the distribution of firms' size is right skew is well known at least since Gibrat (1931). As a matter of fact, the strong form of Gibrat's law implies that the distribution is lognormal, while contemporary analysis points to the power law as the most accurate statistical description of the empirical distribution. It can be argued, however, that the weak form of Gibrat's law is not inconsistent with a power law distribution of firms' size.

So far, the literature has dealt with (*i*) and (*ii*) as if they were independent stylized facts. In this paper we make three contributions.

- We explore the link between the two, showing that the *power law distribution of firms' size* is at the root of the *Laplace distribution of growth rates* (i.e. we demonstrate that, under very general assumptions, (*i*) implies (*ii*)).

- We demonstrate that the features of *business fluctuations* such as the shifts of the distribution of firms' size over the cycle, the properties of the distribution of individual and aggregate growth rates and many others,[1] are a consequence of (*i*).

- We discuss a model of *financial fragility*, empirically validated through conditioning (Brock, 1999), which generates (*i*).

In our approach the origin of business fluctuations – which is the most important single problem in macroeconomics – can be traced back to the ever changing configuration of the network of heterogeneous interacting firms.[2] A major role in shaping dynamics is played by financial variables. The sequential timing structure of our economy implies that future markets are absent, so that agents have to rely on means of payment – here, bank credit extended to firms – to bridge the gap between agents' decisions and their realization. Highly leveraged (i.e., financially fragile) firms, in turns, are exposed to a high risk of default, that is of going bankrupt. When bankruptcies occur, loans not refunded negatively affect banks' net worth, with banks responding to their worsen financial position by reducing credit supply. The reduction in credit supply impacts on the lending interest rate all other firms have to pay to serve their financial commitments.

---

[1] Among them the properties of the distribution of the cumulative growth rates during expansions and contractions; of the age of exiting firms; of profits and "bad debt". See also Delli Gatti *et al.* (2003a).

[2] Schumpeter (1939) suggested that business cycle scholars should analyze "… how industries and individual firms rise and fall and how their rise and fall affect the aggregates and what we call loosely general business conditions". This approach is reminiscent of Marshall's parallel between the dynamics of the individual firm and the evolution of a tree in the forest.

The practice of combining heterogeneity and interactions is at odds with mainstream macroeconomics which reduces the analysis of the aggregate to that of a single *representative* agent and which is unable, by construction, to explain non-normal distributions, scaling behavior or the occurrence of large aggregate fluctuations as a consequence of small idiosyncratic shocks.

While the industrial organization literature has explored at large the evidences (*i*) and (*ii*) at least since the 1950s,[3] scarce attention has been paid so far to establishing a link to business cycle theory, mainly because mainstream macroeconomics lacks the adequate conceptual and analytical tools to accomplish such an endeavor. Its reductionist methodology implies that in order to understand the working of a system, one has to focus on the working of each single element. Assuming that elements are similar and do not interact – i.e. the economy is completely described by a representative agent -- the dynamics of the aggregate replicate the dynamics of the sub-unit. This assumption requires that every element is in equilibrium.

If the system is far from equilibrium, self-organizing phenomena and a state of self-organized criticality (SOC) may occur. According to the notion of SOC (Bak, 1996; Nørrelykke and Bak, 2002), scaling emerges because the sub-units of a system are heterogeneous and interact, and this leads to a *critical state* without any attractive points or states[4]. The occurrence of a power law may be read as a symptom of self organizing processes at work. A notable example of this approach applied to macroeconomics is the inventory and production model developed by Bak *et al.* (1993).

---

[3] For a review of the debate on the shape of the firms' size distribution sprung up during the 1950s and 60s, see the monograph by Steindl (1965).

[4] In the SOC literature the concept of equilibrium is borrowed from *statistical mechanics* and is very different from that of mainstream economics. In fact, equilibrium results from the balance of actions of a large number of many interacting particles.

Alternatively, and in some sense in a way more germane to the economics discourse, power laws can be generated by models based on scale free growth processes. The basic idea can be traced back to the well-known Simon's model (Simon, 1955), where the Gibrat's law of proportional effects is combined with an entry process to obtain a Levy distribution for firms' size. Furthermore, recent work by physicists (e.g. Marsili and Zhang, 1998; Amaral *et al.*, 1998) has shown that, by extending the heterogeneity of the system's components implied in Simon's scheme to account for direct or indirect interactions among units, power laws emerge naturally and, most notably, without the disturbing asymptotic implications of the original Simon's model or of its modern successors, like the one by Gabaix (1999).[5]

It is worthwhile to stress that, regardless of the modeling strategy one chooses, the adoption of the scaling perspective in economics implies rejecting the very definition of a representative agent because the dynamics of the system originate from the interaction of heterogeneous agents. We believe that, in order to grasp the empirical evidence and provide a coherent framework, economists have to adopt a methodological approach based on heterogeneous interacting agents (HIA) (see Delli Gatti *et al.*, 2003a, for an example).

A step in this direction is the agent-based modeling strategy, which is increasingly applied also in economics. Agent-based models have been developed to study the interaction of many heterogeneous agents. In a sense they are based on new microfoundations. The relevance and reliability of these microfoundations are grounded in the empirical evidence they can account for. Microfoundations can be defined as sound if they produce an economic behavior coherent with the empirical evidence, not necessarily with some optimizing principle.

The main shortcuts and the analytical weaknesses of the *representative agent* hypothesis are dealt with in Section 2, where the *pros* of the *HIA* methodology are also illustrated. Section 3 is devoted to discuss the relationship between power law distributions and the business cycle. Section 4 introduces the model (whose simulative details can be found in the Appendix) and discusses its dynamical properties. Section 5 deals with simulation results, focusing on their consistency with the empirical evidence on some business cycle facts (5.1), and on the conditional properties of the model (5.2). Section 6 concludes.

---

[5] Discussions on this point can be found in Krugman (1996) and Blank and Solomon (2000).

## 2. Representative Agent, heterogeneous agents and interaction: a methodological digression

Mainstream economics is based on *reductionism*, i.e. the methodology of classical mechanics. Such a view is coherent if the law of large numbers holds true, i.e.:

- the functional relationships among variables are linear; and,
- there is no direct interaction among agents.

Since non-linearities are pervasive, mainstream economics generally adopts the trick of linearizing functional relationships. Moreover agents are supposed to be all alike and not to interact. Therefore an economic system can be conceptualized as consisting of several identical and isolated components, each one being a representative agent (RA). The *optimal* aggregate solution can be obtained by means of a simple summation of the choices made by each optimizing agent. Moreover, if the aggregate is the sum of its constitutive elements, its dynamics cannot but be identical to that of each single unit.

The ubiquitous RA, however, is at odds with the empirical evidence (Stoker, 1993)[6], is a major problem in the foundation of general equilibrium theory (Kirman, 1992)[7,8] and is not coherent with many econometric investigations and

---

[6] A modeling strategy based on the representative agent is not able, by construction, to reproduce the persistent heterogeneity of economic agents, captured by the skewed distribution of several industrial variables, such as firms' size, growth rates etc. Stoker (1993) reviews the empirical literature at disaggregated level which shows that heterogeneity matters since there are systematic individual differences in economic behavior. Moreover, as Axtell (1999, p.41) claims: "… given the power law character of actual firms' size distribution, it would seem that equilibrium theories of the firm […] will never be able to grasp this essential empirical regularity."

[7] According to Hildenbrand and Kirman (1988, p. 239): "… There are no assumptions on […] isolated individuals which will give us the properties of aggregate behavior which we need to obtain uniqueness and stability. Thus we are reduced to making assumptions at the aggregate level, which cannot be justified, by the usual individualistic assumptions. This problem is usually avoided in the macroeconomic literature by assuming that the economy behaves like an individual. Such an assumption cannot be justified in the context of the standard economic model and the way to solve the problem may involve rethinking the very basis on which this model is founded." This long quotation summarizes the conclusion drawn by Arrow (1951), Sonnenschein (1972), and

tools (Lippi and Forni, 1997)[9]. All in all, we may say that macroeconomics (and macroeconometrics) still lacks sound microfoundations.

The search for *natural laws* in economics does not necessarily require the adoption of the reductionist paradigm. Scaling phenomena and power law distributions are a case in point. If a scaling behavior exists, then the search for universality can be pushed very far. Physicists have shown that scaling laws are generated by a system with strong HIA (Marsili and Chang, 1999; Amaral *et al.*, 1998) and therefore are incompatible with reductionism. As a consequence, the occurrence of scaling laws in economics is incompatible with mainstream economics. The macroscopic pattern (consisting of a multitude of heterogeneous interacting units) works as a unified whole independent of the dynamical process governing its individual components. The idea that systems which consist of a large number of interacting agents *generates* universal, or scaling, laws that do not depend on microscopic details is now popular in statistical physics and is gaining *momentum* in economics as well.

The quantum revolution of last century radically changed the perspective in contemporary physics, leading to a widespread rejection of reductionism. According to the holistic approach, the aggregate is different from the sum of its components because of the interaction of particles. The properties of the sub-units are not intrinsic but can be grasped only analyzing the behavior of the aggregate as a whole. The concept of equilibrium is therefore different from that of mainstream economics. The equilibrium of a system does not require any more that every element is in equilibrium, but rather that the aggregate is quasi-stable,

---

Mantel (1976) on the lack of theoretical foundations of the proposition according to which the properties of an aggregate function reflect those of the individual components.

[8] In General Equilibrium theory one can put all the heterogeneity s/he likes, but no direct interaction among agents. Grossman and Stiglitz (1980) has shown that in this case one cannot have any sort of informational perfection. If information is not perfect markets cannot be efficient. Market failure leads to agents' interaction and to coordination failures, emerging properties of aggregate behavior, and to a pathological nature of business fluctuations.

[9] If agents are heterogeneous, some standard procedures (e.g. cointegration, Granger-causality, impulse-response functions of structural VARs) loose their significance. Moreover, neglecting heterogeneity in aggregate equations generates spurious evidence of dynamic structure. The difficulty of testing aggregate models based on the RA hypothesis, i.e. to impose aggregate regularity at the individual level, has been long pointed out by Lewbel (1989) and Kirman (1992) with no impact on the mainstream (a notable exceptions is Carroll, 2001).

i.e. in "… a state of macroeconomic equilibrium maintained by a large number of transitions in opposite directions" (Feller, 1957, p. 356). [10]

In terms of a power law distribution, it means that firms are located along a curve whose coefficient is stable and the intercept changes very slowly over time.[11] This is due to the fact that the data generating process is random: in terms of the *state of a process* we may say that the transition from one state to another is affected by *chance* as well by agents' *systematic actions*.[12]

In the model of Section 4, output fluctuations are due to: 1) a random process on current revenues as a consequence of imperfect information on actual prices; 2) systematic interactions among agents. The distribution is quasi-stable over relatively long periods because it represents "… slowly changing, age-dependent characteristics of a population which ages and renews itself only gradually" (Steindl, 1965, p.142). This means that, since firms are born small, their growth takes time and mortality decreases with age and size, the slow change of the distribution comes as a consequence. In a nutshell: the distribution is stable, or quasi-stable, because the dissipative force of the process (here, the Gibrat's law) produces a tendency to a growing dispersion, which is counteracted by a stabilizing force (i.e., the burden of debt commitments and the associated risk of bankruptcy).

Moreover, distributions are interconnected. The population is characterized by a joint distribution of several variables (in our model: equity, capital, debt, age, equity ratio), which is completely inconsistent with the RA framework.[13] The change of firms' distribution (and the business cycle itself) has to be analyzed in terms of changes of the joint distribution of the population.[14]

---

[10] Moreover agents' choice should not necessarily be an equilibrium one, derived from their optimizing behavior, because agents' interaction generates self-organizing solutions. It follows from this that one should not analyze the individual problem in isolation from the others (a game *against nature*) but rather the interconnections among HIAs.

[11] Stability of the slope through time is a quite standard result in the empirical literature on Pareto's law (see e.g., the work by C. Gini, J. Steindl and H. Simon). Quite nicely, Steindl (1965, p.143) defines the Pareto coefficient "… a sediment of growth over a long time".

[12] The biased behavior of this *random process* helps to explain the systematic differences (*asymmetries*) between expansions and contractions found in the empirical evidence. Gaffeo *et al.* (2003) have found systematic differences of the Pareto exponents during expansions and contractions.

[13] Marshall once said that there are several representative agents, meaning that the RA framework is a statistical tool, whose average values are actually representative of several distribution.

## 3. Gibrat, Pareto, Laplace: the statistical analysis of industrial dynamics

In this section we set up and discuss some statistical foundations for the results that will be presented in Section 5. As already pointed out, in our HIA framework the analysis of business cycles is intertwined with the properties and evolution of the firms' size distribution, which in turn is empirically well approximated by a power law. A well known object mainly in physics and biology, the power law distribution has been originally derived more than a hundred years ago by Vilfredo Pareto, who argued that the distribution of personal incomes above a certain threshold $y_0$ follows a heavy-tailed distribution (Pareto, 1897). In particular, he showed that the probability to observe an income $Y$ greater than or equal to $y$ is proportional to a power of $y$:

$$\Pr(Y \geq y) \propto y^{-\alpha} \tag{1}$$

with $\alpha$ close to 1.5. This fact immediately started to baffle scholars given that, under the reasonable assumption that the rates of growth of income brackets are only moderately correlated, the Central Limit Theorem implies that the income distribution should be lognormal. This conundrum regarding the limiting distribution of multiplicative stochastic processes became explicit about 30 years later in industrial economics, thanks to the pioneering work of Gibrat (1931). Gibrat claimed that – the law of proportional effects holding true, i.e. that the growth rate of each firm is independent of its size (Gibrat's law in weak form) – the distribution of firms' size must be right skewed. He went even further, arguing that, if the rates of growth are only moderately correlated, such distribution must be a member of the log-normal family (Gibrat's law in strong form). In a nutshell, the size (measured by output, capital stock or number of employed workers) of the

---

[14] A recession, e.g., is more likely when firms are relatively young, small and leveraged. The RA framework not only is inconsistent with the evidence (*i*) but it also misses and outguesses any dynamical properties of the actual systems (Forni and Lippi, 1997).

*i*-th firm $K_{iT}$ in period *T* is defined as $K_{iT} = K_{iT-1}(1+g_{iT})$, where $g_{iT}$ is the rate of growth. Taking the log of both sides and solving back recursively from a time 0 size $K_{i0}$, it is straightforward to obtain $\log K_{iT} \cong \sum_{t=1}^{T} g_{it} + \log K_{i0}$. Assuming that the growth rates are identically independently distributed, the distribution of the log of firms' size tends asymptotically – i.e. for *t* approaching infinity – to the lognormal distribution. Notice that from Gibrat's analysis one would expect the distribution of firms' growth rates to be normal.

Recent empirical research (Axtell, 2001; Gaffeo *et al.*, 2003)[15] has shown, however, that the distribution of firms' size follows a Zipf or power law[16] instead of a lognormal distribution. Moreover, Stanley *et al.* (1996) and Amaral *et al.* (1997) have found that the growth rate of firms' output $\zeta_i$ follows, instead of a normal distribution, a Laplace distribution:

$$L(\zeta_i, b) = \frac{b}{2} \exp(-b\zeta_i) \tag{2}$$

where *b* > 0 is the scale parameter.

To explain this puzzle, the literature has followed two lines of research. The first one is a-theoretical and focuses only on the statistical properties of the link between the distribution of the state variable (firms' size) and that of the rates of change. For instance, Reed (2001) shows that independent rates of change do not generate a lognormal distribution of firms' size if the time of observation of firms' variables is not deterministic but is itself a random variable following approximately an exponential distribution. In this case, even if Gibrat's law holds at the individual level, firms' variables will converge to a double Pareto distribution.

The second line of research – to which the model described in the following section belongs - stresses the importance of non-price interactions among firms hit by multiplicative shocks, hence building on the framework put forward by Herbert Simon and his co-authors during the 1950s and 60s (Ijiri and Simon, 1977). As a matter of example, Bottazzi and Secchi (2003) obtain a Laplace distribution of firms' growth rates within Simon's model, just relaxing the assumption of independence of firms' growth rates.[17]

---

[15] But see also Ijiri and Simon (1977).
[16] To be precise, the *Zipf's law* is the discrete counterpart of the Pareto continuous distribution (power law). It links the probability to observe the dimension of a social or natural phenomenon (firms, cities, earthquakes, words in a text, etc.) with rank greater than, say, *r* with the cumulative frequency. In case of firms' size the scale parameter is equal to 1. In other words, the cumulative frequency is proportional to the inverse of firm's dimension: $\Pr(K_{it} \geq \kappa) \propto \kappa^{-1}$.

This result can be further generalized, in that a Laplace distribution for growth rates can be derived as soon as the size of the state variable under scrutiny is Pareto distributed. As discussed in Palestrini (2003), one can start from the definition of the growth rate as the log-difference of the state variable's levels, so that the proof consists in showing that: 1) the logarithm of a Pareto random variable follows an exponential distribution; and 2) the difference of two exponential random variables becomes a Laplace distribution.

Proposition 1) may be proved using the monotonic property of the logarithmic function and the rule of transformation of random variables. Assuming that $x$ follows a Pareto distribution with parameter $\alpha$ it is possible to derive the probability distribution of $Y = \log(X)$:

$$\begin{aligned} \Pr(Y \geq y) &= \Pr(\log(x) \geq y) \\ &= \Pr(x \geq \exp(y)) \propto (\exp(y))^{-\alpha} \\ &= \exp(-\alpha y) \end{aligned} \quad (3)$$

that is, an exponential distribution with parameter $\alpha$. In the case of independent exponential variables, proposition 2) can be proved – albeit in a more cumbersome way – using the convolution theorem and its relation with the characteristic function. Furthermore, one can show that the causal relationship between Pareto (for levels) and Laplace (for growth rates) distributions holds also in the general case, but the shape of the Laplace distribution of growth rates changes over the business cycles (Palestrini, 2003). In particular, the scale parameter *b* decreases during expansions and increases during recessions.

---

[17] In principle these results can induce the reader to reject the strong version of Gibrat's law. After all, this law claims that the distribution of the levels (firms' size measured in output or capital units) is lognormal – while the empirical analysis points to Zipf's law - and the distribution of growth rates is normal – while it seems to be Laplace. As a matter of fact, things are not that simple. The idea according to which Gibrat's law has to be fully discarded is wrong, given that in the recent literature a weak version seems to hold, in which growth rates seem to be independent at least in mean. In fact, Lee *et al.* (1998) show that the variance of growth rates depends negatively on firm's size. The implications of the strong version of Gibrat's law are not necessarily true in the weak version. Fujiwara *et al.* (2003) have shown, in fact, that if the distribution is characterized by *time-reversal symmetry* – i.e. the joint probability distribution of two consecutive years is symmetric in its arguments $P_{12}(x_1,x_2) = P_{12}(x_2,x_1)$ – the weak version of Gibrat's law can yield a power law of firms' size. Hence power law and Gibrat's law (weak version) are not necessarily inconsistent.

The main message looming large from the statistical results discussed so far is that the scaling approach to business fluctuations derives in the first place from the levels of state variables being distributed as a power law. Thus, the basic question to be answered is whether scale invariance for state variables' levels is a general feature of economic systems or not. From this viewpoint, it emerges that power law probability functions arise endogenously in economics basically for two reasons: 1) the lack of a characteristic scale in empirical and theoretical economics, implying that the occurrence of either rare or frequent events (i.e., sizes) is governed by the same law (Zajdenweber, 1997); 2) a power law behavior in the tail(s) of a distribution is a feature of a family of distributions known as *Lévy-stable distributions*. Due to a generalization of the central limit theorem (Gnedenko and Kolmogorov, 1954), the sum of a large number of identical and independent random variables has a probability density function characterized by a four-parameter characteristic function *M(t)*, which in logarithm reads as:

$$\ln M(t) = \begin{cases} i\mu t - \gamma |t|^\alpha [1 - i\beta \, \text{sgn}(t) \tan(\pi\gamma/2)], & \alpha \neq 1 \\ i\mu t - \gamma |t|[1 + i\beta \, \text{sgn}(t) \frac{2}{\pi} \ln(\gamma)], & \alpha = 1 \end{cases} \qquad (4)$$

where $0 < \alpha \leq 2$ is the stability parameter, $\gamma$ is a positive scale factor, $\mu$ is a real number, and $\beta$ is the skewness parameter ranging from -1 to 1. Expanding *M(t)* in Taylor series it is possible to show (e.g., Mantegna and Stanley, 2000) that the probability density function of a Lévy random variable *x* is, in the tails, proportional to $|x|^{-(1+\alpha)}$. Lévy distributions are stable under convolution, meaning that the sum of *N* identically independently distributed (*iid*) $\alpha$-stable Lévy variables is also Lévy distributed, with the same stability parameter $\alpha$.

When $\alpha = 2$, $\beta = 0$ and $\gamma = \frac{\sigma^2}{2}$, the distribution is Gaussian with mean $\mu$ and variance $\sigma^2$. The Gaussian family is the only member of the Lévy class for which the variance exists. The presence of second moments implies that, if disturbances hitting firms are only idiosyncratic ones, aggregate fluctuations disappear as the number of firms *N* grows large. In fact, without aggregate shocks the variance of the average output of *N* firms is less than the maximum variance of firms' output, say $\frac{\sigma_{\max}^2}{N}$, a quantity that, for *N* going to infinity, vanishes. On the contrary, stable distributions with *α* < 2 do not need aggregate shocks to generate aggregate fluctuations. [18], [19]

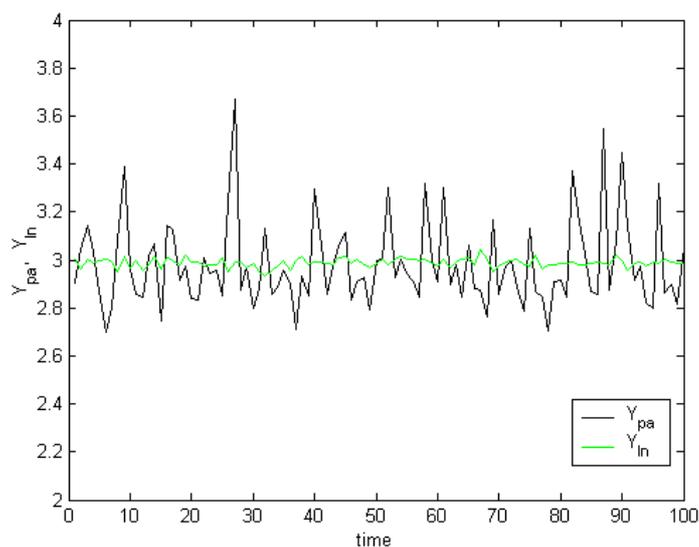

**Figure 1**. Comparison between two simulated economies, inhabited by 10000 firms each. In the first economy (black line) agents' size distribution is Pareto with location parameters *k* = 1 and stability parameter *α* = 1.5. In the second economy (grey line) agents' size distribution is lognormal, with the same mean (i.e., 3) and same estimated variance at *t*=0 (i.e., 10.4) of the other economy. The plot describes the two time series of agents' average output, $Y_{pa}$ (the time evolution of the mean of the Pareto distributed firms) and $Y_{ln}$ (the time evolution of the mean of the lognormal distributed firms).

The difference between the two situations may be well described by the example depicted in Figure 1, where we report the average output time path of two economies identical but for the shape of their firms' size distributions. In the first economy firms are Pareto distributed, whereas in the second one the distribution of firm's size is lognormal. Output is assumed to be proportional to size. Time series have been obtained by averaging from samples extracted from the two distributions at any time period $t$.

---

[18] The first author conjecturing it explicitly has been Mandelbrot (1960).
[19] From this viewpoint, in real world normality (in terms of Gaussian distributed variables) might just be a special case.

The time evolution of the average output shows almost no aggregate fluctuations for the lognormal economy, but large fluctuations in the Pareto economy even in the absence of aggregate shocks. In particular, the variance of the average aggregate output in the Pareto case is one order of magnitude greater than the variance of the lognormal case. Put differently, stable Pareto-Lévy distributions are good candidates to explain aggregate large fluctuations in time periods characterized by small aggregate shocks.

In what follows, we build on the HIA framework developed in Gallegati *et al.* (2003) and Delli Gatti *et al.* (2003a) to put at work all the notions we surveyed in this section, by modeling an economy characterized by aggregate scaling behaviors due to multiplicative idiosyncratic shocks and interactions among firms.

# 4. Financial fragility and business fluctuations: a model

Consider a sequential economy,[20] with time running discretely in periods $t = 1$, 2,…, populated by many firms and banks. Two markets are opened in each period: the market for an homogenous produced good, and the market for credit. As in the *levered aggregate supply* class of models first developed by Greenwald and Stiglitz (1990, 1993), our model is fully supply-determined,[21] in the sense that firms can sell all the output they (optimally) decide to produce.

---

[20] Recall that in a sequential economy (Hahn, 1982) spot markets open at given dates, while future markets do not operate.

[21] Two scenarios are consistent with this assumption. In the *equilibrium* scenario, aggregate demand accommodates supply, i.e. households and firms absorb all the output produced by the latter and the goods market is always in equilibrium. In this scenario, aggregate investment must be equal to the sum of retained profits and households' saving. As we will see, both investment and retained profit are determined in the model, so that we have to assume that households' saving adjusts in order to fill the gap between the two. In the *disequilibrium* scenario, aggregate demand does not (necessarily) accommodate supply, so that the goods market is generally not in equilibrium. In this case, the difference between aggregate investment on the one hand, and the sum of profit and households' saving on the other must be assumed to take the form of involuntary inventories decumulation.

Due to informational imperfections on the equity market, firms can raise funds only on the credit market. The demand for credit is related to investment expenditure, which is therefore dependent on banks' interest rates. Total credit supply, in turn, is a multiple of the banks' equity base, which is negatively affected as insolvent borrowing firms go bankrupt. As we will discuss below, this mean-field interaction provides a mechanism to create long-range inter-temporal correlations capable to amplify and propagate idiosyncratic shocks.

## 4.1 Firms

At any time period $t$, the supply side of the economy consists of finitely many competitive firms indexed with $i = 1, \ldots, N_t$, each one located on an island. The total number of firms (hence, islands) $N_t$ depends on $t$ because of endogenous entry and exit processes to be described below. Let the $i$-th firm uses capital ($K_{it}$) as the only input to produce a homogeneous output ($Y_{it}$) by means of a linear production technology, $Y_{it} = \phi K_{it}$. Capital productivity ($\phi$) is constant and uniform across firms, and the capital stock never depreciates.

The demand for goods in each island is affected by an *iid* idiosyncratic real shock. Since arbitrage opportunities across islands are imperfect, the individual selling price in the $i$-th island is the random outcome of a market process around the average market price of output $P_t$, according to the law $P_{it} = u_{it}P_t$, with expected value $E(u_{it}) = 1$ and finite variance.

By assumption, firms are fully rationed on the equity market, so that the only external source of finance at their disposal is credit. The balance sheet identity implies that firms can finance their capital stock by recurring either to net worth ($A_{it}$) or to bank loans ($L_{it}$), $K_{it} = A_{it} + L_{it}$. Under the assumption that firms and banks sign long-term contractual relationships, at each $t$ debt commitments in real terms for the $i$-th firm are $r_{it}L_{it}$, where $r_{it}$ is the real interest rate.[22] If, for the sake of simplicity, the latter is also the real return on net worth, each firm incurs financing costs equal to $r_{it}(L_{it} + A_{it}) = r_{it}K_{it}$. Total variable costs proportional to financing costs[23], $gr_{it}K_{it}$, with $g > 1$. Therefore, profit in real terms ($\pi_{it}$) is:

---

[22] It follows that the credit lines periodically extended by the bank to each firm are based on a mortgaged debt contract.

[23] As a matter of example, one can think of retooling and adjustment costs to be sustained each time the production process starts.

$$\pi_{it} = u_{it}Y_{it} - gr_{it}K_{it} = (u_{it}\phi - gr_{it})K_{it}. \tag{5}$$

and expected profit is $E(\pi_{it}) = (\phi - gr_{it})K_{it}$.

In this economy, firms may go bankrupt as soon as their net worth becomes negative, that is $A_{it} < 0$. The law of motion of $A_{it}$ is:

$$A_{it} = A_{it-1} + \pi_{it}, \tag{6}$$

that is, net worth in previous period plus (minus) profits (losses). Making use of (5) and (6), it follows that the bankruptcy state occurs whenever:

$$u_{it} = \frac{1}{\phi}\left(gr_{it} - \frac{A_{it-1}}{K_{it}}\right) \equiv \bar{u}_{it}. \tag{7}$$

As in Greenwald and Stiglitz (1990, 1993), the probability of bankruptcy ($\Pr^f$) is incorporated directly into the firm's profit function because going bankrupt costs, and such a cost is increasing in the firm's output. Assuming for expositional convenience that $u_{it}$ is uniformly distributed on the support (0,2), and that bankruptcy costs are quadratic, $C^f = cY_{it}^2$ with $c > 0$, the objective function takes the form:[24]

$$\Gamma_{it} = (\phi - gr_{it})K_{it} - \frac{\phi c}{2}\left(gr_{it}K_{it}^2 - A_{it-1}K_{it}\right). \tag{8}$$

From the first order condition, the optimal capital stock is:

$$K_{it}^d = \frac{\phi - gr_{it}}{c\phi gr_{it}} + \frac{A_{it-1}}{2gr_{it}}. \tag{9}$$

Thus, the desired capital stock in $t$ is decreasing (non-linearly) with the interest rate and it increases linearly with financial robustness, as proxied by the $t$–1 net worth. Time period $t$ desired investment is simply the difference between the desired capital stock and the capital stock inherited from the previous period, $I_{it} = K_{it}^d - K_{it-1}$. To finance it, the $i$-th firm recurs to retained profits and, if needed, to new mortgaged debt, $I_{it} = \pi_{it-1} + \Delta L_{it}$,[25] where $\Delta L_{it} = L_{it} - L_{it-1}$. Making use of (4), the demand for credit is given by:

$$L_{it}^d = \frac{(\phi - gr_{it})}{c\phi gr_{it}} - \pi_{it-1} + \left(\frac{1 - 2gr_{it}}{2gr_{it}}\right) A_{it-1}. \tag{10}$$

---

[24] For this program to be well defined and the second term on the l.h.s. of (4) to be interpreted as an expected bankruptcy cost, $g$ should be such that the condition $A_{it-1} < gr_{it}K_{it}$ holds.

## 4.2 The banking sector

We model the banking sector in terms of the reduced form from the working of an oligopolistic industry. The balance sheet of the banking sector is $L_t^s = E_t + D_t$, with $L_t$ being total credit supply, $E_t$ the banks' equity base and $D_t$ deposits which, in this framework, are determined as a residual. To determine the aggregate level of credit supply, we assume that banks are subject to a prudential rule set up by a regulatory body such that $L_t^s = E_{t-1}/\nu$, where the risk coefficient $\nu$ is constant. Hence, the healthier are banks from a financial viewpoint, the higher is the aggregate credit supply (Hubbard *et al.*, 2002).

Credit is allotted to each individual firm $i$ on the basis of the mortgage it offers, which is proportional to its size, and to the amount of cash available to serve debt[26] according to the rule:

$$L_{it}^s = \lambda L_s \frac{K_{it-1}}{K_{t-1}} + (1 - \lambda) L_s \frac{A_{it-1}}{A_{t-1}} \tag{11}$$

---

[25] A word of caution is in order here. The law of motion of the net worth (4) seems to imply that the correct time at which profit had to be taken into account in deriving the demand for credit should be time period $t$. In fact, the timing structure of the model is such that when deciding how much to borrow from banks, firms do not have received any time $t$ revenues yet. Hence, at the beginning of time period $t$ the only internal finance they can count on are inherited equity and time $t$–1 profits.
[26] For evidence on the effects exerted by firms' size and credit worthiness on banks' loan policies see e.g. Strahan (1999).

with $K_{t-1} = \sum_{i=1}^{N_{t-1}} K_{it-1}$, $A_{t-1} = \sum_{i=1}^{N_{t-1}} A_{it-1}$, and $0 < \lambda < 1$. The equilibrium interest rate for the *i*-th firm is determined as credit demand (10) equals credit supply (11), that is:

$$r_{it} = \frac{2 + A_{it-1}}{2cg\left(\frac{1}{\phi c} + \pi_{it-1} + A_{it-1}\right) + 2cgL_s\left[\lambda \kappa_{it-1} + (1-\lambda)\alpha_{it-1}\right]}. \tag{12}$$

where $\kappa_{it-1}$ and $\alpha_{it-1}$ are the ratios of individual to total capital and net worth, respectively.

Under the assumption that the returns on the banks' equity are given by the average of lending interest rates $\bar{r}_t$, while deposits are remunerated with the borrowing rate $r_t^A$, the banks' profit ($\pi_t^B$) is given by:

$$\pi_t^B = \sum_{i \in N_t} r_{it} L_{it}^s - \bar{r}_t\left[(1-\omega)D_{t-1} + E_{t-1}\right] \tag{13}$$

with $\frac{1}{1-\omega}$ being the spread between lending and borrowing rates. Note that $\omega$, which in what follows will be treated parametrically, captures the degree of competition in the banking sector: the higher is $\omega$, the higher is the interests' spread which, in turn, increases with a higher monopolistic power of banks.

When a firm goes bankrupt, $K_{it} < L_{it}$. In this case, the banking sector as a whole registers a loss equal to the difference between the total amount of credit supplied up to time period *t* and the relative mortgage, $B_{it} = L_{it} - K_{it} = -A_{it}$, where $A_{it} < 0$ if firm *i* belong to the set of bankrupt firms $\Omega_t$. Let us call $B_{it}$ *bad debt*. The banking sector's equity base evolves according to the law of motion:

$$E_t = \pi_t^B + E_{t-1} - \sum_{i \in \Omega_{t-1}} B_{t-1}. \tag{14}$$

Through the banking sector's equity base law of motion, idiosyncratic real disturbances leading to a bankruptcy have systemic consequences: an increase of bad debt forces the aggregate credit supply shifting to the left, thus raising the financial costs due to a higher interest rate, *ceteris paribus*. Furthermore, the distribution of firms' net worth influences the average lending interest rate, which in turn affects the bank's profit and, eventually, credit supply. Thus, firms dynamically affect each other through indirect interactions. In particular, interactions are global and independent of any topological space, and they occur through a field variable, which in our case is the banking sector's balance sheet (Aoki, 1996).

Interactions, if strong enough, allow the system to escape from the property of square root scaling for sums of *iid* shocks due to the Central Limit Theorem.[27] It is well known from statistic theory (e.g, Resnik, 1987) that as $N$ grows large, independence of idiosyncratic disturbances implies that the volatility of the system decays with the square root of size, leading to a power law distribution with exponent $\beta = -0.5$. If *distant* agents are sufficiently correlated through interactions, in turn, aggregate volatility decays more slowly, according to a power law with exponent $\beta < -0.5$. The empirical evidence reported in Amaral *et al*. (1997) for companies and in Canning *et al*. (1998) for countries goes precisely in this direction.

## 4.3 Firms' demography

Recent empirical work has shown that firms entering and exiting markets contribute almost as much to employment and macroeconomic fluctuations as firms continuing their activity (e.g., Davis *et al*., 1996). Hence, any theory of business fluctuations should pay particular attention to the way entry and exit of firms are modeled.[28]

In our framework, exits are endogenously determined as financially fragile firms go bankrupt, that is as their net worth becomes negative. Besides making the total output to shrink, exits cause the equity of the banking sector – and, in turn, aggregate credit – to go down. As discussed above, this mean field interaction in terms of a *bank effect* (Hubbard *at al*., 2002) amplifies and propagates idiosyncratic shocks all over the economy.

---

[27] This is another way to explicate the arguments reported in section 3.
[28] Delli Gatti *et al.* (2003b) provide an extensive analysis on the relationship between entries and exits and aggregate fluctuations in a model very similar to this one.

As regards entries, the literature has suggested models ranging from exogenously defined purely stochastic processes (Winter *et al.*, 1997), to models where entry is endogenous in that the number of entrants depends on expected profit opportunities (Hopenhayn, 1992). Alas, the available evidence has been so far inconclusive. Caves (1998), for instance, claims that the only firm points are that entrants are in general largely unsure about the probability of prospective success, and that entries does not occur at a unique sector-specific optimal size.

Our modeling strategy aims at capturing these facts by means of a mechanism in which a probabilistic process is affected by prospective performance, and entries can take place at different sizes. First, the number of new entrants ($N_t^{entry}$) is obtained by multiplying a constant $\overline{N} > 1$ to a probability which depends negatively on the average lending interest rate:

$$N_t^{entry} = \overline{N} \Pr(entry) = \frac{\overline{N}}{1 + \exp[d(\overline{r}_{it-1} - e)]} \tag{15}$$

where *d* and *e* are constants. The higher is the interest rate, the higher are financial commitments, and the lower are expected profits, with entries being lower in number. Second, entrants' size in terms of their capital stock is drawn from a uniform distribution centered around the mode of the size distribution of incumbent firms, each entrant being endowed with an equity ratio ($a_{it} = \dfrac{A_{it}}{K_{it}}$) equal to the mode of the equity base distribution of incumbents.

## 4.4 Long-run dynamics

In order to understand the long-run – i.e., growth – properties of our economy, it is convenient to consider a deterministic version of the model. Indeed, abstracting from uncertainty means getting rid of heterogeneity, so that we can easily keep track of the dynamic behavior of a representative firm. If the interest rate is assumed constant, from (5), (6) and (9) it turns out that the law of motion of the net worth is:

$$A_t = \left(1 + \frac{\phi - gr}{2gr}\right) A_{t-1} + \frac{(\phi - gr)^2}{c\phi gr} \tag{16}$$

The solution of this first order difference equation returns the steady state gross growth rate of the economy, $\frac{1}{2}\left[\frac{\phi}{gr} - 1\right]$, which implies positive growth whenever $(\phi - gr) > 0$: whenever the return to capital is higher than its cost, the economy is characterized by endogenous growth. This result is far from surprising as soon as we note that in our model the production function exhibits constant returns to the only input that can be accumulated, which is the same engine of growth as in the well-known *AK* endogenous growth model developed by Rebelo (1991).

This analogy can be further extended to appreciate the special role played by credit in our economy. First, recall that in the Rebelo's model the steady-state growth rate depends positively on the saving rate. In our partial equilibrium analysis savings are implicitly defined as the difference between investment and retained profits,[29] so that at each time period *t* total savings are equal to banks' loans. Indeed, changes in the banking regulatory regime or in the competitive pressure in the banking sector end up affecting the equilibrium lending interest rate, and through it the long-run growth rate.

## 5. Simulation results

The complexity of the model directs the analysis of its high-frequency properties towards computer simulation techniques. Figures 2 and 3 exhibit the evolution of an artificial economy lasting 1000 time periods, implemented using the framework analyzed in the previous section with a starting number of 10000 firms.[30] In particular, in Figure 2 we show the time path of the logarithm of total output, whereas in Figure 3 it is drawn its volatility expressed in terms of output's growth rates.

---

[29] See the *equilibrium* scenario depicted in note 5.
[30] In the following, every simulation were conducted by means of *Swarm*, an agent-based software developed at the Santa Fe Institute to implement artificial economies. Interested readers can find it at the web site: www.swarm.org.

From Figure 2 it emerges that our stochastic economy, buffeted with *iid* idiosyncratic disturbances only, is characterized by sizable aggregate fluctuations; that its growth process displays a broken-trend behavior[31] (Perron, 1987); and that *Great Depressions* (e.g., the one during the simulation time period 855-880) can suddenly punctuate its time path, due to bankruptcies of great firms that origin remarkable impacts on the business cycle *via* the financial sector (Gabaix, 2003). The output series possesses an autocorrelation parameter equal to 0.99. Interestingly, before large downturns our model seems to exhibit a common pattern: starting from a constant growth trend, the economy gains momentum with accelerating growth and increasing volatility, to subsequently move into a deep recession.

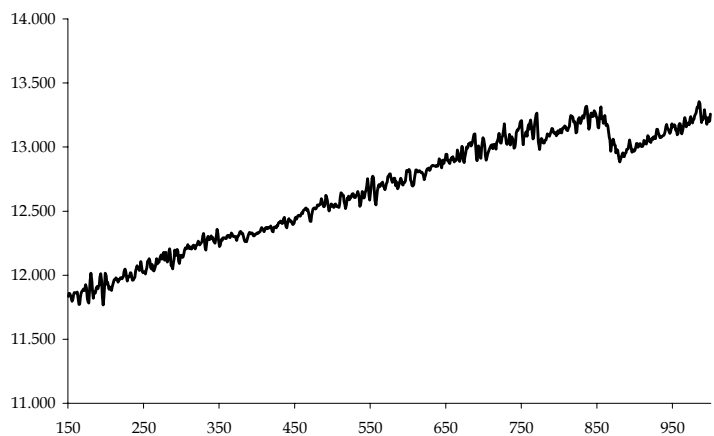

**Figure 2**. Logarithm of the aggregate output. The first 150 periods have been deleted to get rid of transients.

Furthermore, as shown in Figure 3, fluctuations as measured by output's growth rates are characterized by cluster volatility, a well known phenomenon mostly in the financial market literature due to the heavy tails character of asset

---

[31] For instance, the average growth rate goes from 0.19% in periods 150-350, to 0.25% in periods 351-780, to 0.37% in periods 780-855, to 0.31% in periods 880-1000. Yearly average growth rates more close to reality could be obtained in this model through a more careful calibration exercise. Given that our main interest is in business fluctuations, however, we leave this undertaking to future research.

returns' distributions (Cont *et al.*, 1997). The growth rates' standard deviation is 0.0289.

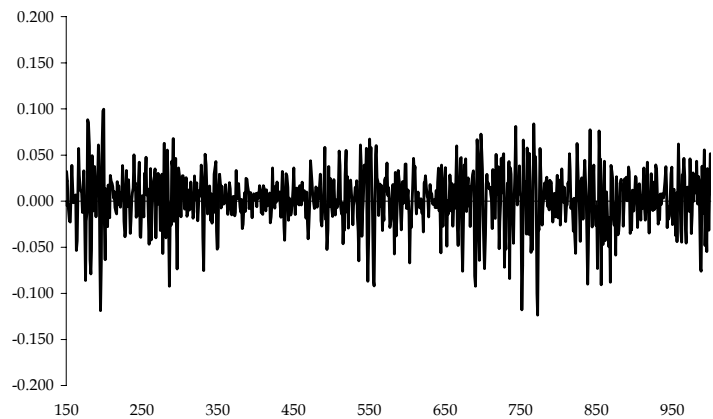

**Figure 3**. Growth rates of aggregate output.

## 5.1 Firms' size and growth rates distributions

In Figure 4 we report the Zipf plot for firm sizes recorded at simulation time period 1000. In agreement with recent empirical results (Axtell, 2001) the firms' size distribution is skewed and it follows a power law. Furthermore, the scaling exponents recorded ($\alpha$ = 1.11) are consistent with what found in real data (Gaffeo *et al.*, 2003). As widely shown in the complexity literature, the emergence of such a distribution is deeply correlated with the hypothesis of interaction of heterogeneous agents that is at the root of the model. More specifically, the interaction among units buffeted with multiplicative *iid* shocks leads the system's dynamics to a complex critical state in which no attractive point or state emerges. In terms of business fluctuations, it means that there is not a single and determinate equilibrium, but a non-stable state emerges after each recessive or expansive episode.

The firms' size distribution tends to shift to the right during growing phases, while during recessions the estimated stability parameter $\alpha$ decreases. In fact, during expansions greater firms tend to grow faster than smaller ones, causing a higher slope of the interpolating line if compared with the situation observed during recessions. On the contrary, bankruptcies of great firms during downturns cause a more equal distribution of the size distribution. Once again, this is precisely what observed in real data (Gaffeo *et al.*, 2003).

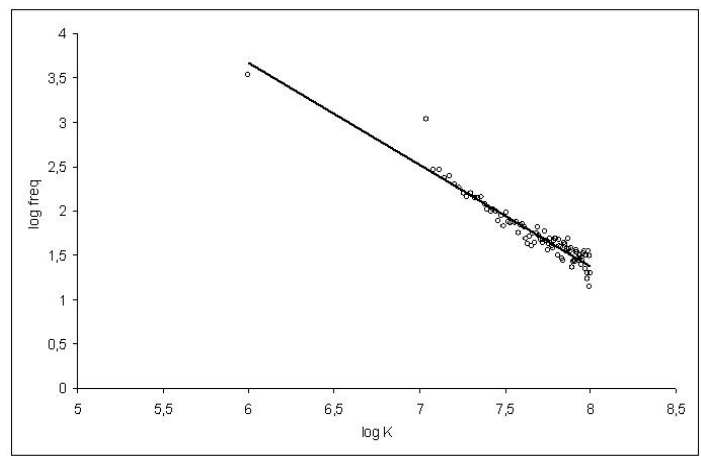

**Figure 4**. Zipf plot of firm sizes.

Stanley *et al.* (1996) and Bottazzi and Secchi (2003), among others, find that the growth rates of firms are generally well fitted by a Laplace (or double exponential) distribution. As discussed in Section 3, such a finding can be shown to derive from firms' size being distributed as a power law. In fact, simulated data for firms' growth rates, reported in Figure 5, are well approximated by a (asymmetric) Laplace distribution.

In another stimulating paper, Lee *et al.* (1998) show that binned growth rates for firms and countries' GDPs settle on the same regression line in a log-log plot. If analyzed from a complex perspective, this result signals the presence of *self similarity*[32], i.e. the behavior of greatest units (countries) reproduces the behavior of smaller units (firms), possibly corrected by a scale factor (Durlauf, 2003). As shown in Figure 5, where we plot the distribution of the growth rates of aggregate output, this feature has been recorded in our model as well. The difference of parameters between the firms' growth rate and the aggregate output growth rates distributions is sensible, in our simulations, to the modeling choice for the production function, though we do not have at this stage any analytical result to prove it.

---

[32] According to Sornette (2000, p.94), self-similarity occurs when "… arbitrary sub-parts are statistically similar to the whole, provided a suitable magnification is performed along all directions".

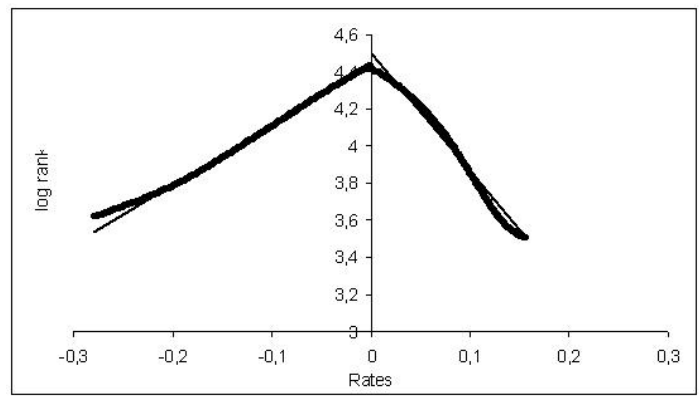

**Figure 5**. Distribution of firms growth rates.

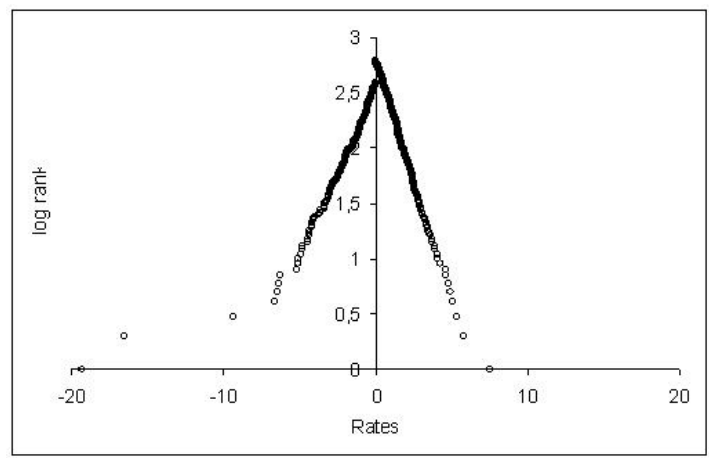

**Figure 6**. growth rates of aggregate output.

The model is capable to display several other striking similarities with observable facts[33]. In particular: 1) the frequency of firms' exits seems to be well approximated by an exponential function of firms' age (Steindl, 1965; Fujiwara, 2003); 2) bad debt, that is the amount of unpaid loans due to bankruptcies extended by the banking sector, follows a stretched exponential distribution (Delli Gatti *et al.*, 2003d); 3) profits are power law distributed, and exhibit time reversal symmetry (Fujiwara, 2003); 4) expansions and recessions, measured as trough-to-peak and peak-to-trough of the GDP growth rates time series, are distributed as a Weibull (Di Guilmi *et al.*, 2003); 5) the rate of return on the capital $\left(\dfrac{\pi_i}{K_i}\right)$ and the equity ratio $a_i$ are positively correlated; 6) a higher equity ratio is associated with a lower volatility of profits, the last two facts being consistent with the evidence one can obtained by analyzing the sample of firms in Bureau van Dijk's AMADEUS, a commercially available data set.

## 5.2 Conditional distributions

In this subsection we address a typical aggregation issue, known as the *mixture* problem, which is likely to negatively affect the reliability of results as soon as scaling plots are taken into account (Brock, 1999; Durlauf, 2003). Roughly speaking, the mixture problem asserts that, when aggregating economic units with different behaviors, it is possible to observe marginal distributions with heavy tails even though conditional distributions do not possess such a property. In other terms, a power law may appear simply because heterogeneous units governed by different stochastic processes are erroneously mixed, instead of signaling the invariant properties of a unique Lévy-stable underlying stochastic process. In fact, if the latter is the case one should observe the same scaling behavior independently of which conditioned sub-sample is considered. In fact, the mixture problem may be present in our work, since the model described in Section 4 implies different behaviors according to the financial position of firms, as well as differently aged firms.

---

[33] Details can be found in companion papers (Delli Gatti *et al.*, 2003b; 2003c).

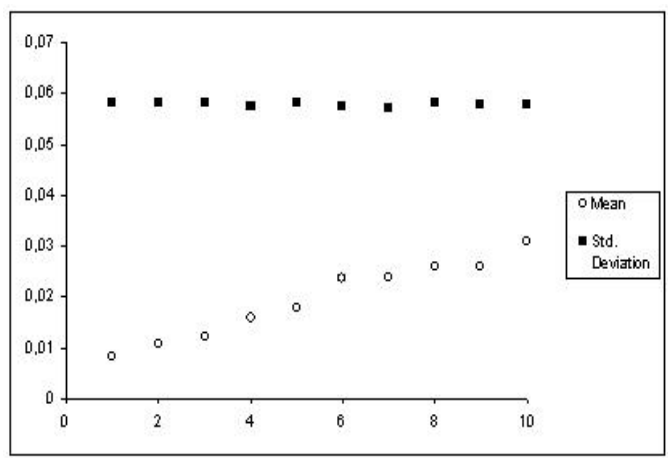

**Figure 8**. Profit distribution conditioned on the equity ratio, with firms grouped in 10 bins.

To understand which variable is likely to be most suitable for conditioning, we start considering one of the basic hypothesis at the root of the model mechanics, that is the fact that an heavy indebted firm is forced to use a large amount of its revenues to pay for its financial commitments, instead of using it for real investments. In other terms, a high leverage is likely to reduce the profitability index. The analysis of the relationship between profit rate and equity ratio, conducted by means of nonparametric regression[34], returns an upward sloping trend (Figure 8) as one would expect from the theoretical model, and in line with what recorded for empirical data. Furthermore, simulations show that the rate of profit distribution shifts to the right when conditioning on the equity ratio[35], and that the probability to fail does not depend on the size but only on the financial position[36]. This is important for the analysis to follow, suggesting that to address the mixture problem it is sufficient to compute firms' distributions conditional on the equity ratio.

---

[34] We use a kernel density estimation (Härdle, 1990), with a Gaussian kernel.
[35] The shape of the conditioned profit distributions depends on the assumption one makes on the distribution of idiosyncratic real shocks. We made several trials, to conclude that the best approximation to what observed in real data could be obtained by forcing the relative price shocks to be normally distributed. Nevertheless, none of the model's properties discussed in the main text are qualitatively affected by this modeling choice.
[36] This finding could be further conceived by recalling that firs' exists are exponentially distributed, and that the exponential distribution possesses the well known property of being memoryless.

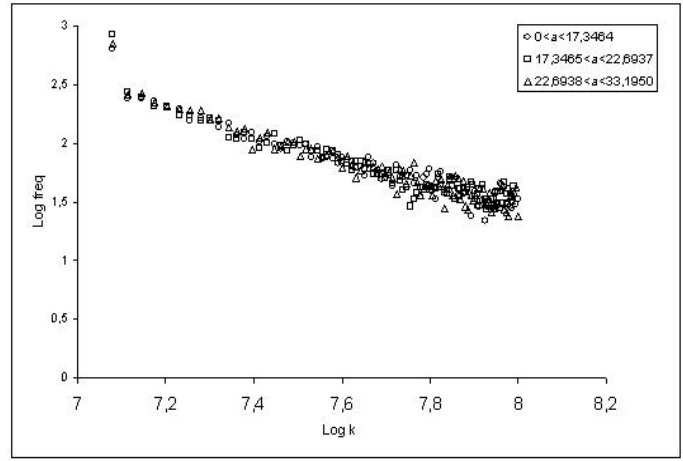

**Figure 9**. Zipf plot of firms dimension sorted by equity ratio (a).

Hence, the power law behavior of the firms' size distribution is analyzed partitioning the [0,1] interval, to which *a* belongs, in several bins (chosen according to a percentile allotment). Figure 9 shows that data from different distributions conditioned on the equity partition [0, 0.1734], (0.1734, 0.2269], (0.2269, 0.3319][37] collapse on the same interpolating line, a clear sign of self-similarity, thus signaling that the unconditional distribution of firms' size is likely to display a scaling behavior because of its true nature and not due to spurious mixing.

As one would expect by combining this finding and the results summarized in Section 3, the level of the equity ratio seems not to have any influence on the relative growth of firms, since the conditional distributions of growth rates, sorted in bins according to their financial position, invariably collapse on the same curve.

To summarize, of the two forms of heterogeneity in the model – i.e., firms' financial position and age – the one that really matters in firms' behavior is the former, here measured by the equity ratio *a*. The analysis above shows that the scaling and the self-similarity properties – phenomena which suggest complex

---

[37] A fourth bin, i.e. the partition (0.3319, 1], has been excluded from the plot due to a lack of sufficient observations.

behaviors – do not depend on the aggregation of different economic units but it is an intrinsic property of an economic system with interacting units buffeted with idiosyncratic multiplicative shocks.

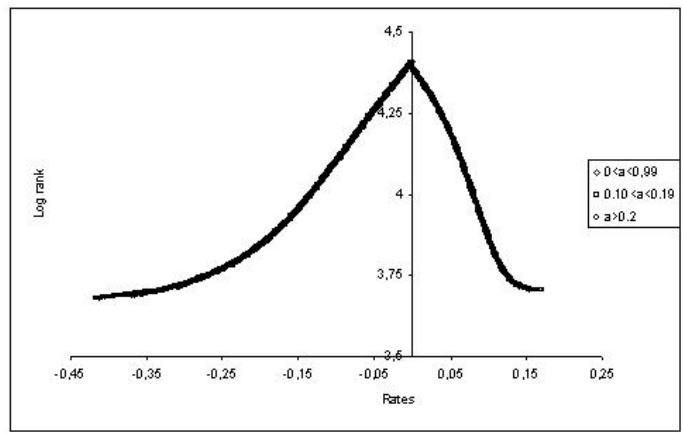

**Figure 10**. Distributions of firms' growth rates partitioned according to their equity ratio.

## 6. Conclusions

Scaling phenomena and power law distributions are rather unfamiliar concepts for scholars interested in business cycle theory, regardless of the fact that these objects have been studied in economics since a long time. The reason for this neglectfulness should be looked for in the reductionism methodology which has so far permeated modern macroeconomics. Our position is that the reductionism paradigm is not only theoretically unsatisfying, but it can also be falsified as soon as proper new stylized facts are isolated. Concepts and methods inspired from physics have revealed particularly useful in detecting new facts and guiding theory formation. This paper aims at popularize the scaling approach to business fluctuations, by discussing some scaling-based ideas involved in viewing the macroeconomy as a complex system composed of a large number of heterogeneous interacting agents (HIAs).

In particular, we present a simple agent-based model of the *levered aggregate supply* class developed by Greenwald and Stiglitz (1990, 1993), whose core is the interaction of heterogeneous financially fragile firms and a banking sector. In order to grasp the empirical evidence we adopt a methodological approach based on agent-based simulations of a system with HIAs. In our framework, the origin of business fluctuations can be traced back to the ever changing configuration of the network of heterogeneous interacting firms.

Simulations of the model replicate surprisingly well an impressive set of stylized facts, particularly two well known *universal laws*: *i*) the distribution of firms' size (measured by the capital stock) is skewed and described by a power law; *ii*) the distribution of the rates of change of aggregate and firms' output follow a similar Laplace distribution. So far, the literature has dealt with stylized facts (*i*) and (*ii*) as if they were independent. We have discussed as that *power law distribution of firms' size* lays at the root of the *Laplace distribution of growth rates.*

The model can be extended in a number of ways to take into account, among other things, the role of aggregate demand, different degrees of market power on the goods and credit markets, technological change, policy variables, learning processes, etc. Our conjecture, however, is that the empirical validation of more complex models will be due to the basic ingredients already present in the benchmark framework: the power law distribution of firms' size and then associated Laplace distribution of growth rates which in turn can be traced back to the changing financial conditions of firms and banks.

# Appendix

In this appendix we briefly describe the assumptions and procedures we followed to simulate the model. A simulation is completely described by the parameter values, the initial conditions and the rules to be iterated period after period. First of all, we set the parameter values and the initial conditions for state variables needed to start the simulation. There parameters of the model are relative to the firm, bank and the entry process.

For the firm we have:
- the productivity of capital $\phi$,
- the parameter of the bankruptcy cost equation $c$,
- the firm's equity-loan ratio $\alpha$,
- the variable cost parameter $g$

For the bank:
- the mark down on interest rate $\omega$,
- the weight the bank gives to the capital in allotting the credit supply $\lambda$.

For the entry process:
- the location parameter $d$,
- the sensitivity parameter $e$,
- the size parameter $\bar{N}$.

They are set as follows: $\phi = 0.1$ ; $c = 1$ ; $\alpha = 0.08$, $g = 1.1$ ; $\omega = 0.002$, $\lambda = 0.3$, $d = 100$, $e = 0.1$. $\bar{N}$ must be set according to the initial number of firms (see below). The first step of the simulation occurs at time $t=1$. To perform calculations in period 1 for each firm we must set initial conditions for firms' capital, the equity base, profit and bad debt. We chose the following values $K_{i0} = 100$, $A_{i0} = 20$, $L_{i0} = 80$, $\pi_{i0} = 0$, $B_{i0} = 0$. We run simulations for several values of initial firms and for different number of iterations. In the simulation we report on Section 5 the initial number of firms was set to 10000 and the number of iterations to 1000. Given the initial number of firms, we set $\bar{N} = 180$. The main loop is described in the following algorithm.

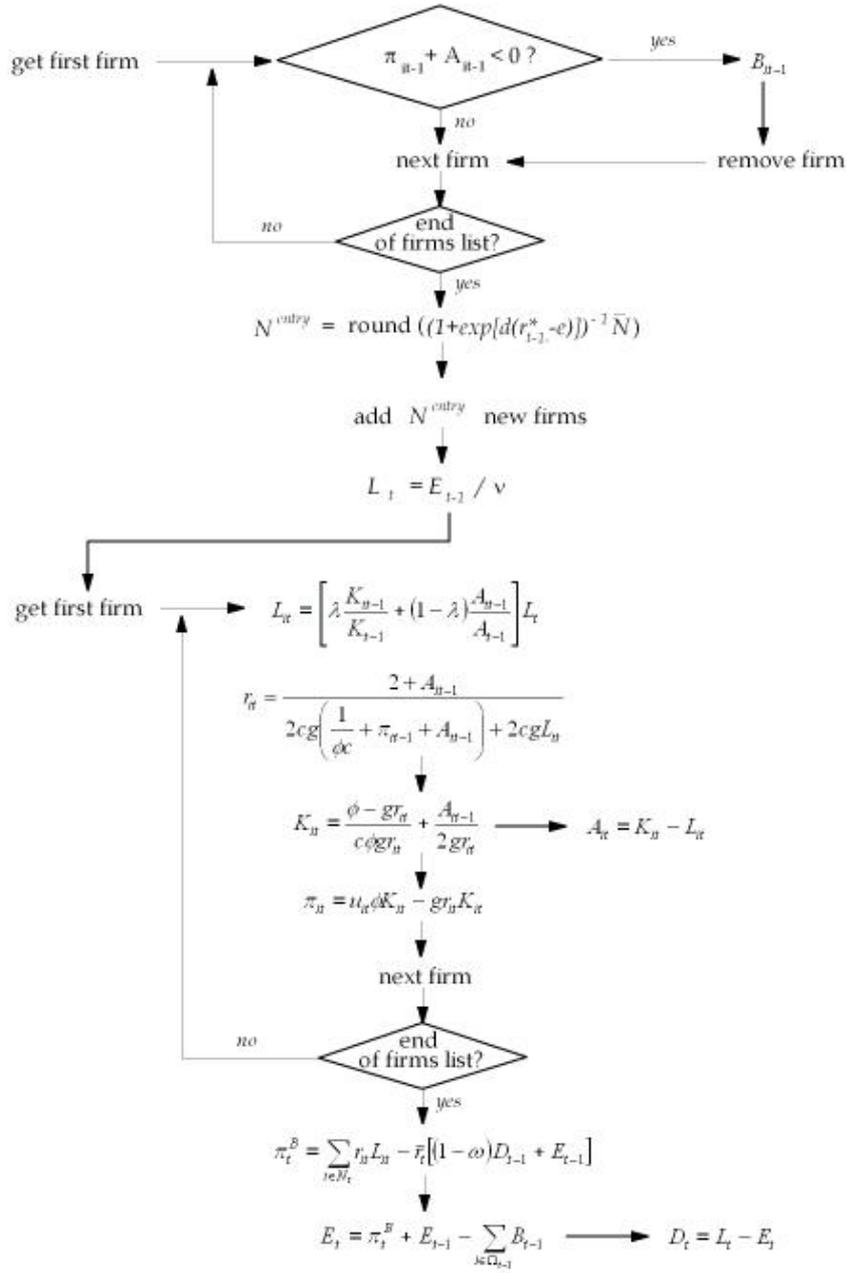